\documentclass[12pt]{article}

\author{Yu.~M.~Zinoviev
       \thanks{E-mail address: Yurii.Zinoviev@ihep.ru} \\[0.5cm]
        {\it Institute for High Energy Physics} \\
        {\it of National Research Center "Kurchatov Institute"} \\
        {\it Protvino, Moscow Region, 142280, Russia}}

\title{On massive higher spin  supermultiplets in $d=3$}

\date{}

\begin{document}

\maketitle

\begin{abstract}
In this paper, using a frame-like gauge invariant formulation of the
massive higher spin bosons and fermions, we develop a direct
construction of the completely off-shell cubic vertices describing an
interaction of the massless gravitino with the massive higher spin
supermultiplets. To achieve the invariance under the local
supersymmetry we introduce all necessary supertransformations (both
for the physical as well as for the auxiliary fields) and thus all the
supercurrents constructed are conserved on-shell. As an illustration
of the technique used we present some lower superspin examples and then
we consider the arbitrary superspin. We also check that the whole
construction is completely consistent with all bosonic and
fermionic gauge symmetries of the fields entering the supermultiplets.
\end{abstract}

\thispagestyle{empty}
\newpage
\setcounter{page}{1}

\section{Introduction}

Surprisingly, a task of explicit construction for the massive higher
spin supermultiplets appears to be much more complicated than a
similar task for the massless ones. As far as the global supersymmetry
is concerned, the procedure seems to be pretty straightforward: one
has to take two famous Singh-Hagen Lagrangians \cite{SH74,SH74a} for
massive boson and fermion which differ in spins by 1/2 and find
supertransformations leaving the sum of these two Lagrangian
invariant. However, starting already with vector supermultiplet, one
has to introduce some complicated higher derivative corrections to the
supertransformations. Moreover, the higher the spins one tries to
consider, the more derivatives one has to introduce. And this makes
the task almost intractable\footnote{Only recently the superfield
analogue for the Singh-Hagen procedure has been realized for the
massive supermultiplets with half-integer superspins \cite{Kou20}.}.

In the metric-like formalism a general solution was suggested in
\cite{Zin07a} on the base of the gauge invariant description of
massive higher spin bosons \cite{Zin01} and fermions \cite{Met06}. The
main idea was to construct massive supermultiplet using an appropriate
set of the massless ones exactly in the same way as the gauge
invariant description of massive fields can be obtained starting with
the appropriate set of the massless ones. With the development of the
frame-like formalism for the massive higher spin fields
\cite{Zin08b,PV10,KhZ19}, this approach becomes even more effective
and allowed us to construct not only massive supermultiplets
\cite{BKhSZ19}, but the supermultiplets containing partially-massless
fields \cite{BKhSZ19a} and infinite spin fields \cite{BKhSZ19b} as
well.

As in many other aspects, three dimensions appear to be special also
on the properties of the massive higher spin supermultiplets.
In-particular, there exist supermultiplets containing so-called
topologically massive higher spin fields \cite{KT16,KP18}. At the same
time the construction of the massive supermultiplets similar to the
four-dimensional one was successfully realized \cite{BSZ15,BSZ16}
(see \cite{BSZ17a} for review) using the gauge invariant formulation
for massive bosons and fermions specially adopted to three dimensions 
\cite{BSZ12a,BSZ14a}. 

Let us turn now to the local supersymmetry. Again the task seems to be
straightforward. Having in our disposal the global
supertransformations we just have to make them local. Then the sum of
the two free Lagrangian ceases to be invariant but this must be
compensated by the interaction with the massless gravitino. However in
so doing one faces the ambiguities related with a lot of possible field
redefinitions\footnote{For the general discussions on the field
redefinitions specific to the gauge invariant formulation of the
massive fields see \cite{BDGT18}.}. One of the ways to construct cubic
vertices describing an interaction of the massless gravitino and the
massive supermultiplet is to use the so-called Fradkin-Vasiliev
formalism \cite{FV87,FV87a}. In three dimensions this formalism does
not work for the massless fields but does work for the massive ones
\cite{Zin16}. And indeed this approach was realized in
\cite{BSZ17,BSZ17a}. However, FV-formalism involves  the so-called
extra fields (which do not enter the free Lagrangians but appears only
in  the interactions) making the formalism essentially to be
on-shell. Moreover, generally it produces terms having much more
derivatives that is absolutely necessary. This number of derivatives
can be diminished again by the field redefinitions but it requires a
lot of work\footnote{As an illustration, see what happens in the case
of massless fields and massless supermultiplets in $d=4$
\cite{KhZ20a,KhZ20b}.}.

Our aim in this work is a direct construction of the completely
off-shell cubic vertices describing interactions of the massless
gravitino and the massive arbitrary superspin multiplets having
minimal number of derivatives possible. To make these vertices to be
gauge invariant we introduce all necessary supertransformations both
for the physical as well as for the auxiliary fields. This in turn
guarantee that the corresponding supercurrents  are conserved on-shell.
One of the technical problems here is that lower spin components are
zero-forms, while all higher spin ones are one-forms. So one faces the
situations when some zero-form must transform into one-form. There are
two possibilities here. From one hand, one can introduce an inverse
frame allowing to convert world indices into local ones with the risk
to spoil the compactness and elegance of the frame-like formalism. The
other approach, which we follow in this work, is to switch in such
cases to the metric-like multispinor formalism, similar to the ones
used in \cite{KT16,KP18,KhZ22}. Let us stress that all cubic vertices
constructed in our work are written in the pure frame-like formalism,
while the metric-like one is used only to write the
supertransformations for the zero-forms as well as at some
intermediate steps of calculations. 

This paper is organized as follows. In section 2 we provide all
necessary information on the gauge invariant formulation for the
massive higher spins in $d=3$. In section 3 we present a number of the
lower superspin examples mainly to illustrate the usage of the
metric-like formalism for the lower spin components. At last in
section 4 we construct the cubic interaction vertex for the arbitrary
superspin multiplet as well as the complete set of the
supertransformations.

\noindent
{\bf Notation and conventions} We work in the frame-like multispinor
formalism where all fields are one-forms or zero-forms having a set of
completely symmetric local spinor indices $\alpha=1,2$. For coordinate
free description of the flat three dimensional space we use background
frame $e^{\alpha(2)}$ and background Lorentz covariant derivative $D$
such that
\begin{equation}
D \wedge D = 0, \qquad D \wedge e^{\alpha(2)} = 0 
\end{equation}
Besides $e^{\alpha(2)}$, a complete basis of forms contain two-form
$E^{\alpha(2)}$ and three-form $E$ defined as follows
\begin{equation}
e^{\alpha(2)} \wedge e^{\beta(2)} = \epsilon^{\alpha\beta}
E^{\alpha\beta}, \qquad E^{\alpha(2)} \wedge e^{\beta(2)} =
\epsilon^{\alpha\beta} \epsilon^{\alpha\beta} E
\end{equation}
As it was already mentioned, each time as it is convenient we switch
to the metric-like multispinor formalism using the fact that any 
one-form can be transformed into a set of zero-forms (see
\cite{KhZ22} for details), for example
\begin{eqnarray*}
H^{\alpha(2)} &=& e_{\beta(2)} h^{\alpha(2)\beta(2)} + 
e^\alpha{}_\beta h^{\alpha\beta} + e^{\alpha(2)} h \\
\Phi^\alpha &=& e_{\beta(2)} \phi^{\alpha\beta(2)} + e^\alpha{}_\beta
\phi^\beta 
\end{eqnarray*}

\section{Kinematics}

In this section we provide all necessary information on the frame-like
gauge invariant description for the massive higher spin bosons and
fermions. As it was already mentioned in the Introduction, such
formalism plays an essential role in the constriction of the massive
higher spin supermultiplets in any dimensions. 

\subsection{Massive bosons $s \ge 2$}

In this case the frame-like gauge invariant formulation requires
\cite{BSZ12a} a set of one-forms $\Omega^{\alpha(2k)}$,
$H^{\alpha(2k)}$, $1 \le k \le s-1$, $A$ and zero-forms
$B^{\alpha(2)}$, $\pi^{\alpha(2)}$ and $\varphi$. The free Lagrangian
in the flat Minkowski background has the form:
\begin{eqnarray}
{\cal L}_0 &=& \sum_{k=1}^{s-1} (-1)^{k+1} 
[ k \Omega_{\alpha(2k-1)\beta} e^\beta{}_\gamma 
\Omega^{\alpha(2k-1)\gamma} + \Omega_{\alpha(2k)} D H^{\alpha(2k)} ]
\nonumber \\
 && + E B_{\alpha(2)} B^{\alpha(2)} - B_{\alpha(2)} e^{\alpha(2)}
D A - E \pi_{\alpha(2)} \pi^{\alpha(2)} + \pi_{\alpha(2)}
E^{\alpha(2)} D \varphi \nonumber \\
 && + \sum_{k=1}^{s-2} (-1)^{k+1} a_k [ - \frac{(k+2)}{k}
\Omega_{\alpha(2k)\beta(2)} e^{\beta(2)} H^{\alpha(2k)} + 
\Omega_{\alpha(2k)} e_{\beta(2)} H^{\alpha(2k)\beta(2)} ] \nonumber \\
 && + 2a_0 \Omega_{\alpha(2)} e^{\alpha(2)} A - a_0 
H_{\alpha\beta} E^\beta{}_\gamma B^{\alpha\gamma} + 2Ms
\pi_{\alpha(2)} E^{\alpha(2)} A \nonumber \\
 && + \sum_{k=1}^{s-1} (-1)^{k+1} b_k H_{\alpha(2k-1)\beta} 
e^\beta{}_\gamma H^{\alpha(2k-1)\gamma} + \frac{Msa_0}{2}
H_{\alpha(2)} E^{\alpha(2)} \varphi + \frac{3}{2}a_0{}^2 E \varphi^2,
\end{eqnarray}
where
\begin{equation}
a_k{}^2 = \frac{k(s+k+1)(s-k-1)}{2(k+1)(k+2)(2k+3)}M^2, \quad
a_0{}^2 = \frac{(s+1)(s-1)}{3}M^2, \quad 
b_k = \frac{M^2s^2}{4k(k+1)^2}.
\end{equation}
This Lagrangian is invariant under the following local gauge
transformations:
\begin{eqnarray}
\delta \Omega^{\alpha(2k)} &=& D \eta^{\alpha(2k)} + \frac{(k+2)}{k}
a_k e_{\beta(2)} \eta^{\alpha(2k)\beta(2)} 
+ \frac{a_{k-1}}{k(2k-1)} e^{\alpha(2)} \eta^{\alpha(2k-2)} +
\frac{b_k}{k} e^\alpha{}_\beta \xi^{\alpha(2k-1)\beta}, \nonumber \\
\delta H^{\alpha(2k)} &=& D \xi^{\alpha(2k)} + a_k e_{\beta(2)}
\xi^{\alpha(2k)\beta(2)} + \frac{(k+1)a_{k-1}}{k(k-1)(2k-1)}
e^{\alpha(2)} \xi^{\alpha(2k-2)} + e^\alpha{}_\beta 
\eta^{\alpha(2k-1)\beta}, \nonumber \\
\delta \Omega^{\alpha(2)} &=& D \eta^{\alpha(2)} + 3_1
e_{\beta(2)} \eta^{\alpha(2)\beta(2)} + b_1 
e^\alpha{}_\beta \xi^{\alpha\beta}, \nonumber \\
\delta H^{\alpha(2)} &=& D \xi^{\alpha(2)} + e^\alpha{}_\beta
\eta^{\alpha\beta} + a_1 e_{\beta(2)} \xi^{\alpha(2)\beta(2)}
+ 2a_0 e^{\alpha(2)} \xi, \\
\delta B^{\alpha(2)} &=& 2a_0 \eta^{\alpha(2)}, \qquad
\delta A = D \xi + \frac{a_0}{4} e_{\alpha(2)}
\xi^{\alpha(2)}, \nonumber \\
\delta \pi^{\alpha(2)} &=& \frac{Msa_0}{2} \xi^{\alpha(2)}, \qquad
\delta \varphi = - 2Ms \xi. \nonumber
\end{eqnarray}
Lagrangian equations for the one-forms (up to $(-1)^{k+1}$) look like:
\begin{eqnarray}
0 &=& D \Omega^{\alpha(2k)} + \frac{(k+2)}{k}a_k
e_{\beta(2)} \Omega^{\alpha(2k)\beta(2)} + \frac{a_{k-1}}{k(2k-1)}
e^{\alpha(2)} \Omega^{\alpha(2k-2)} + \frac{b_k}{k} e^\alpha{}_\beta
H^{\alpha(2k1)\beta}, \nonumber \\
0 &=& D H^{\alpha(2k)} + a_k e_{\beta(2)}
H^{\alpha(2)\beta(2)} + \frac{(k+1)a_{k-1}}{k(k-1)(2k-1)}
e^{\alpha(2)} H^{\alpha(2k-2)} + e^\alpha{}_\beta 
\Omega^{\alpha(2k-1)\beta}, \nonumber \\
0 &=& D \Omega^{\alpha(2)} + 3a_1 e_{\beta(2)}
\Omega^{\alpha(2)\beta(2)} + b_1 e^\alpha{}_\beta H^{\alpha\beta}
- \frac{a_0}{2} E^\alpha{}_\beta B^{\alpha\beta} 
+ \frac{Msa_0}{2} E^{\alpha(2)} \varphi, \label{bos_eq} \\
0 &=& D H^{\alpha(2)} + e^\alpha{}_\beta
\Omega^{\alpha\beta} + a_1 e_{\beta(2)} H^{\alpha(2)\beta(2)}
+ 2a_0 e^{\alpha(2)} A, \nonumber \\
0 &=& e_{\alpha(2)} D B^{\alpha(2)} - 2a_0 e_{\alpha(2)}
\Omega^{\alpha(2)} + 2Ms E_{\alpha(2)} \pi^{\alpha(2)}, \nonumber
\end{eqnarray}
while equations for the zero-forms $B^{\alpha(2)}$, $\pi^{\alpha(2)}$
and $\varphi$:
\begin{eqnarray}
0 &=& 2E [ B^{\alpha(2)} - 2 D^\alpha{}_\beta A^{\alpha\beta}
- 2a_0 h^{\alpha(2)} ], \nonumber \\
0 &=& 2E [ D^{\alpha(2)} \varphi - \pi^{\alpha(2)} + 2Ms A^{\alpha(2)}
], \\
0 &=& E [ - 2 (D\pi) + 3Msa_0 h + 3a_0{}^2 \varphi ], \nonumber
\end{eqnarray}
where we use the following relations between one-forms and zero-forms:
$$
D = e_{\alpha(2)} D^{\alpha(2)}, \qquad
A = e_{\alpha(2)} A^{\alpha(2)}, \qquad
H^{\alpha(2)} = e_{\beta(2)} h^{\alpha(2)\beta(2)} + e^\alpha{}_\beta
h^{\alpha\beta} + e^{\alpha(2)} h. 
$$

\subsection{Massive vector}

In this case we use metric-like first order formalism with physical
fields $A^{\alpha(2)}$, $\varphi$ and auxiliary fields
$B^{\alpha(2)}$, $\pi^{\alpha(2)}$. The free Lagrangian:
\begin{equation}
{\cal L}_0 = \frac{1}{2} B_{\alpha(2)} B^{\alpha(2)} + 4
B^{\alpha\beta} D_\alpha{}^\gamma A_{\beta\gamma} - \frac{1}{2}
\pi_{\alpha(2)} \pi^{\alpha(2)} - \pi^{\alpha(2)} D_{\alpha(2)}
\varphi + 2M \pi^{\alpha(2)} A_{\alpha(2)} 
\end{equation}
is invariant under the following gauge transformations
\begin{equation}
\delta A^{\alpha(2)} = D^{\alpha(2)} \xi, \qquad \delta \varphi = 2M
\xi.
\end{equation}
The equations for all fields have the form:
\begin{eqnarray}
0 &=& B_{\alpha(2)} + 2 D_\alpha{}^\beta A_{\alpha\beta}, \qquad
0 = 2 D_\alpha{}^\beta B_{\alpha\beta} + 2M \pi_{\alpha(2)}
\nonumber \\
0 &=& - \pi_{\alpha(2)} - D_{\alpha(2)} \varphi + 2M A_{\alpha(2)},
\qquad 0 = (D\pi). 
\end{eqnarray}

\subsection{Massive scalar}

For the scalar we also use the metric-like first order formalism with
the Lagrangian
\begin{equation}
{\cal L}_0 = - \frac{1}{2} \pi_{\alpha(2)} \pi^{\alpha(2)} -
\pi_{\alpha(2)} D^{\alpha(2)} \varphi - \frac{M^2}{4} \varphi^2,
\end{equation}
and the corresponding equations:
\begin{equation}
0 = - \pi_{\alpha(2)} - D_{\alpha(2)} \varphi, \qquad
0 = (D\pi) - \frac{M^2}{2} \varphi.
\end{equation}

\subsection{Massive fermions $s \ge \frac{3}{2}$}

The frame-like gauge invariant formulation for the massive fermions
with spin $s+1/2$ requires \cite{BSZ14a} a set of one-forms 
$\Phi^{\alpha(2k+1)}$, $0 \le k \le s-1$ and a zero-form 
$\rho^\alpha$. The free Lagrangian has the form
\begin{eqnarray}
\frac{1}{i} {\cal L}_0 &=& \sum_{k=0}^{s-1} (-1)^{k+1}
[ \frac{1}{2} \Phi_{\alpha(2k+1)} D \Phi^{\alpha(2k+1)}]
+ \frac{1}{2} \rho_\alpha E^\alpha{}_\beta D \rho^\beta \nonumber \\
 && + \sum_{k=1}^{s-1} (-1)^{k+1} c_k \Phi_{\alpha(2k-1)\beta(2)}
e^{\beta(2)} \Phi^{\alpha(2k-1)} + c_0 \Phi_\alpha E^\alpha{}_\beta
\rho^\beta \nonumber \\
 && + \sum_{k=0}^{s-1} (-1)^{k+1} \frac{d_k}{2} \Phi_{\alpha(2k)\beta}
e^\beta{}_\gamma \Phi^{\alpha(2k)\gamma} - \frac{3d_0}{2} E
\rho_\alpha \rho^\alpha,
\end{eqnarray}
where 
\begin{equation}
d_k = \frac{(2s+1)}{(2k+3)}M, \qquad
c_k{}^2 = \frac{(s+k+1)(s-k)}{2(k+1)(2k+1)}M^2, \qquad
c_0{}^2 = 2s(s+1)M^2. 
\end{equation}
This Lagrangian is invariant under the following gauge
transformations: 
\begin{eqnarray}
\delta_0 \Phi^{\alpha(2k+1)} &=& D \zeta^{\alpha(2k+1)} +
\frac{d_k}{(2k+1)} e^\alpha{}_\beta \zeta^{\alpha(2k)\beta} \nonumber
\\
 && + \frac{c_k}{k(2k+1)} e^{\alpha(2)} \zeta^{\alpha(2k-1)}
+ c_{k+1} e_{\beta(2)} \zeta^{\alpha(2k+1)\beta(2)}, \\
\delta_0 \rho^\alpha &=& c_0 \zeta^\alpha. \nonumber
\end{eqnarray}
Lagrangian equations for the one-forms (up to $(-1)^{k+1}$) look like
\begin{eqnarray}
0 &=& D \Phi^{\alpha(2k+1)} + \frac{d_k}{(2k+1)} e^\alpha{}_\beta
\Phi^{\alpha(2k)\beta} \nonumber  \\
 && + \frac{c_k}{k(2k+1)} e^{\alpha(2)} \Phi^{\alpha(2k-1)} 
+ c_{k+1} e_{\beta(2)} \Phi^{\alpha(2k+1)\beta(2)}, \label{ferm_eq} \\
0 &=& D \Phi^\alpha + d_0 e^\alpha{}_\beta \Phi^\beta
+ c_1 e_{\beta(2)} \Phi^{\alpha\beta(2)} - c_0 E^\alpha{}_\beta
\rho^\beta,  \nonumber
\end{eqnarray}
while equation for the zero-form $\rho^\alpha$
\begin{equation}
0 = E [ 2 D^\alpha{}_\beta \rho^\beta + 3c_0 \phi^\alpha - 3d_0
\rho^\alpha]. 
\end{equation}
Note that the zero-form $\phi^\alpha$ appears in the decomposition
$$
\Phi^\alpha = e_{\beta(2)} \phi^{\alpha\beta(2)} + e^{\alpha(2)}
\phi^\alpha.
$$

\subsection{Massive spinor}

Here the free Lagrangian is simply
\begin{equation}
\frac{1}{i}{\cal L}_0 = \frac{1}{2} \rho_\alpha D^\alpha{}_\beta
\rho^\beta - \frac{M}{4} \rho_\alpha \rho^\alpha,
\end{equation}
with the corresponding equation:
\begin{equation}
0 = D^\alpha{}_\beta \rho^\beta - \frac{M}{2} \rho^\alpha. 
\end{equation} 

\section{Lower superspins}

In this section we consider some lower superspin examples, mainly to
illustrate the usage of the metric-like formalism for the zero form
components.

\subsection{Superspin $Y = 0$}

We begin with the sum of the free Lagrangians for the spinor and the
scalar:
\begin{equation}
{\cal L}_0 = \frac{1}{2} \rho_\alpha D^\alpha{}_\beta \rho^\beta
- \frac{M_0}{4} \rho_\alpha \rho^\alpha - \frac{1}{2} \pi_{\alpha(2)}
\pi^{\alpha(2)} - \pi_{\alpha(2)} D^{\alpha(2)} \varphi 
- \frac{M^2}{4} \varphi^2,
\end{equation}
with the corresponding equations:
\begin{equation}
0 = - D_\alpha{}^\beta \rho_\beta - \frac{M_0}{2} \rho_\alpha, \qquad
0 = - \pi_{\alpha(2)} - D_{\alpha(2)} \varphi, \qquad
0 = (D\pi) - \frac{M^2}{2} \varphi.
\end{equation}
Now let us consider the following ansatz for the cubic vertex
describing an interaction for such massive supermultiplet with the
massless gravitino:
\begin{equation}
{\cal L}_1 = \Psi_\alpha [ g_1 (E\pi) \rho^\alpha + g_2 
e^{\alpha\beta} D \varphi \rho_\beta + g_3 \varphi
E^{\alpha\beta} \rho_\beta ]. 
\end{equation}
Calculating variations under the local supertransformations we obtain:
\begin{eqnarray*}
\delta_0 {\cal L}_1 &=& \zeta_\alpha [ - g_1 E_{\beta(2)} D
\pi^{\beta(2)} \rho^\alpha - g_1 E_{\beta(2)} \pi^{\beta(2)}
D \rho^\alpha - g_2 e^{\alpha\beta} D \varphi D \rho_\beta \\
 && \quad - g_3 E^{\alpha\beta} D \varphi \rho_\beta - g_3
E^{\alpha\beta} \varphi D \rho_\beta ]. 
\end{eqnarray*}
Switching to the metric-like formalism and omitting three form $E$
as a common multiplier we get:
\begin{eqnarray*}
\delta_0 {\cal L}_1 &=& - 2g_1 \zeta_\alpha (D\pi) \phi^\alpha - 2g_1
\zeta_\alpha \pi_{\beta(2)} D^{\beta(2)} \phi^\alpha \\
 && + 4g_2 \zeta_\alpha (D^{\alpha\gamma} \varphi D^\beta{}_\gamma +
D^{\beta\gamma} \varphi D^\alpha{}_\gamma) \phi_\beta \\
 && - 2g_3 \zeta_\alpha D^{\alpha\beta} \varphi \phi_\beta
- 2g_3 \zeta_\alpha \varphi D^{\alpha\beta} \phi_\beta. 
\end{eqnarray*}
To compensate for these variations we introduce the following
supertransformations for the bosonic fields:
\begin{eqnarray}
\delta_1 \pi^{\alpha(2)} &=& - 2g_2 (\zeta^\beta D^\alpha{}_\beta
\rho^\alpha + \zeta^\alpha D^\alpha{}_\beta \rho^\beta) 
- g_3 \zeta^\alpha \rho^\alpha, \nonumber \\
\delta_1 \varphi &=& 2g_1 \zeta_\alpha \rho^\alpha, \\
\delta_1 \rho^\alpha &=& - 2g_3 \varphi \zeta^\alpha. \nonumber
\end{eqnarray}
They produce
\begin{eqnarray*}
\delta_1 {\cal L}_0 &=& - 2f_2 \zeta_\alpha [ (D^{\alpha(2)} \varphi +
\pi^{\alpha(2)} D_\alpha{}^\beta \rho_\beta + (D^{\beta(2)} \varphi
+ \pi^{\beta(2)}) D^\alpha{}_\beta \rho_\beta ] \\
 && + g_3 \zeta_\alpha [ (D^{\alpha(2)} \varphi + \pi^{\alpha(2)})
\rho_\alpha - 2 \varphi (D^\alpha{}_\beta - \frac{M_0}{2}
\rho^\alpha)] \\
 && + 2g_1 \zeta_\alpha (D\pi) - \frac{M^2}{2} \varphi) \rho^\alpha.
\end{eqnarray*}
We obtain:
\begin{eqnarray*}
\delta_0 {\cal L}_1 + \delta_1 {\cal L}_0 &=& \zeta_\alpha [ - 2g_1
\pi_{\beta(2)} D^{\beta(2)} \rho^\alpha - 2g_2 \pi_{\beta(2)}
D^{\alpha\beta} \rho^\beta + 2g_2 \pi^{\alpha(2)} D_\alpha{}^\beta
\rho_\beta ] \\
 && + \zeta_\alpha [ g_3 \pi^{\alpha(2)} \rho_\alpha 
+ (M_0g_3 - M^2g_1) \varphi \rho^\alpha ]. 
\end{eqnarray*}
Using simple identities
\begin{eqnarray*}
D^{\beta(2)} \rho^\alpha &=& \frac{1}{3} \rho^{\alpha\beta(2)} -
\frac{1}{3} \epsilon^{\alpha\beta} D^\beta{}_\gamma \rho^\gamma \\
D^{\alpha\beta} \rho^\beta &=& \frac{2}{3} \rho^{\alpha\beta(2),}
+ \frac{1}{3} \epsilon^{\alpha\beta} D^\beta{}_\gamma \rho^\gamma,
\end{eqnarray*}
where
$$
\rho^{\alpha(3)} = D^{\alpha(2)} \rho^\alpha,
$$
we obtain
\begin{eqnarray*}
\delta_0 {\cal L}_1 + \delta_1 {\cal L}_0 &=& \zeta_\alpha [ 
- \frac{2(g_1+2_2)}{3} \pi_{\beta(2)} \rho^{\alpha\beta(2)} 
+ \frac{4(g_1 - 4g_2)}{3} \pi^\alpha{}_\beta D^\beta{}_\gamma
\rho^\gamma ] \\
 && + \zeta_\alpha [ g_3 \pi^{\alpha(2)} \rho_\alpha 
+ (M_0g_3 - M^2g_1) \varphi \rho^\alpha ]. 
\end{eqnarray*}
Now we put
\begin{equation}
g_2 = - \frac{g_1}{2} 
\end{equation}
and we finally get
$$
\delta_0 {\cal L}_1 + \delta_1 {\cal L}_0 = \zeta_\alpha [ 4g_1
\pi^\alpha{}_\beta \rho^\beta - 2g_3 \pi^\alpha{}_\beta \rho^\beta
+ (M_0g_2 - M^2g_1) \varphi \rho^\alpha] 
$$
At last, we introduce appropriate supertransformations for the spinor
\begin{equation}
\delta_1 \rho^\alpha = - 4g_1 \pi^{\alpha\beta} \zeta_\beta 
\end{equation}
and this leaves us with
$$
\delta_0 {\cal L}_1 + \delta_1 {\cal L}_0 = \zeta_\alpha
[ 2(M_0g_1 - g_3) \pi^\alpha{}_\beta \rho^\beta
+ (M_0g_3 - M^2g_1) \varphi \rho^\alpha ].
$$
So we must have
\begin{equation}
g_3 = M_0g_1, \qquad M_0{}^2 = M^2
\end{equation}
Thus we obtain the vertex
\begin{equation}
{\cal L}_1 = g_1 \Psi_\alpha [ (E\pi) \rho^\alpha - \frac{1}{2} 
e^{\alpha\beta} D \varphi \rho_\beta + M_0 \varphi E^{\alpha\beta}
\rho_\beta ]
\end{equation}
as well as the following supertransformations
\begin{eqnarray}
\delta \pi^{\alpha(2)} &=& g_1 (\zeta^\beta D^\alpha{}_\beta
\rho^\alpha + \zeta^\alpha D^\alpha{}_\beta \rho^\beta) 
- M_0g_1 \zeta^\alpha \rho^\alpha, \nonumber \\
\delta \varphi &=& 2g_1 \zeta_\alpha \rho^\alpha, \\
\delta \rho^\alpha &=& - 4g_1 \pi^{\alpha\beta} \zeta_\beta 
- 2M_0g_1 \varphi \zeta^\alpha.  \nonumber
\end{eqnarray}

\subsection{Superspin $Y = \frac{1}{2}$}

In our second example we also begin with the sum of the free
Lagrangians for the massive vector and the spinor:
\begin{eqnarray}
{\cal L}_0 &=& \frac{1}{2} B_{\alpha(2)} B^{\alpha(2)} + 4
B^{\alpha\beta} D_\alpha{}^\gamma A_{\beta\gamma} - \frac{1}{2}
\pi_{\alpha(2)} \pi^{\alpha(2)} - \pi^{\alpha(2)} D_{\alpha(2)}
\varphi + 2M \pi^{\alpha(2)} A_{\alpha(2)} \nonumber \\
 && + \frac{1}{2} \rho_\alpha D^\alpha{}_\beta \rho^\beta
- \frac{M_0}{4} \rho_\alpha \rho^\alpha,
\end{eqnarray}
with the complete set of equations:
\begin{eqnarray}
0 &=& B_{\alpha(2)} + 2 D_\alpha{}^\beta A_{\alpha\beta}, \qquad
0 = 2 D_\alpha{}^\beta B_{\alpha\beta} + 2M \pi_{\alpha(2)}, \nonumber
\\
0 &=& - \pi_{\alpha(2)} - D_{\alpha(2)} \varphi + 2M A_{\alpha(2)},
\qquad 0 = (D\pi), \\
0 &=& - D_\alpha{}^\beta \rho_\beta - \frac{M_0}{2} \rho_\alpha.
\nonumber
\end{eqnarray}

In this case a possible ansatz for the cubic vertex is restricted by
the invariance under the gauge transformations of the vector field. We
consider
\begin{eqnarray}
{\cal L}_1 &=& \Psi_\alpha [ g_1 (E^\alpha{}_\beta B^\beta{}_\gamma -
B^\alpha{}_\beta E^\beta{}_\gamma) \rho^\gamma + g_2 DA \rho^\alpha
\nonumber \\
 && \quad + g_3 (E\pi) \rho^\alpha + g_4 e^{\alpha\beta} 
(D \varphi - 2M A) \rho_\beta  ]. 
\end{eqnarray}
Calculating the variations and transforming them into the metric-like
formalism we obtain:
\begin{eqnarray*}
\delta_0 {\cal L}_1 &=& g_1 \zeta_\alpha D^\alpha{}_\beta
B^{\alpha\beta} \phi_\alpha + 2g_1 \zeta_\alpha (B^{\alpha\beta}
D_\beta{}^\gamma \phi_\gamma + B^{\beta\gamma} D_\beta{}^\alpha
\phi_\gamma) \nonumber \\
 && + 4g_2 \zeta_\alpha D_\beta{}^\gamma A_{\beta\gamma} D^{\beta(2)}
\phi^\alpha - 2g_3 \zeta_\alpha (D\pi) \phi^\alpha - 2g_3
\zeta_\alpha \pi_{\beta(2)} D^{\beta(2)} \phi^\alpha \nonumber \\
 && + 2g_4 (D^{\alpha(2)} \varphi - 2MA^{\alpha(2)})
(\zeta_\alpha D^\beta{}_\alpha \phi_\beta + \zeta_\beta 
D^\beta{}_\alpha \phi_\alpha) - 4Mg_4 \zeta_\alpha
D^\alpha{}_\beta A^{\alpha\beta} \phi_\alpha.
\end{eqnarray*}
Now we introduce the following supertransformations for the bosons:
\begin{eqnarray}
\delta_1 B^{\alpha(2)} &=& - 2g_2 \zeta_\beta D^{\alpha(2)} \rho^\beta
- 2Mg_4 \zeta^\alpha \rho^\alpha, \qquad
\delta_1 A_{\alpha(2)} = \frac{g_1}{2} \zeta_\alpha \rho_\alpha,
\nonumber \\
\delta_1 \pi_{\alpha(2)} &=&  2g_4 (\zeta_\alpha D^\beta{}_\alpha
\rho_\beta + \zeta_\beta D^\beta{}_\alpha \rho_\alpha), \qquad
\delta_1 \varphi = 2g_3 \zeta_\alpha \rho^\alpha. 
\end{eqnarray}
They produce
\begin{eqnarray*}
\delta_1 {\cal L}_0 &=& \zeta_\alpha [ - 2g_1 (D^\alpha{}_\beta
B^{\alpha\beta} - M\pi^{\alpha(2)}) \rho_\alpha - 2g_2 \zeta_\alpha
(2D_\beta{}^\gamma A_{\beta\gamma} + B_{\beta(2)}) D^{\beta(2)}
\rho^\alpha + 2g_3 \zeta_\alpha (D\pi) \rho^\alpha \\
 && + \zeta_\alpha [ - 2g_4 (D^{\alpha(2)} \varphi + \pi^{\alpha(2)} -
2M A^{\alpha(2)}) D_\alpha{}^\beta \rho_\beta - 2g_4 (D^{\beta(2)}
\varphi + \pi^{\beta(2)} - 2M A^{\beta(2)}) D^\alpha{}_\beta
\rho_\beta] \\
 && + 4Mg_4 \zeta_\alpha (2D^\alpha{}_\beta A^{\alpha\beta} -
B^{\alpha(2)}) \rho_\alpha. 
\end{eqnarray*}
We obtain
\begin{eqnarray*}
\delta_0 {\cal L}_1 + \delta_1 {\cal L}_0 &=& 
\zeta_\alpha [ \frac{2(g_1-g_2)}{3} B_{\beta(2)}
\rho^{\alpha\beta(2)} - \frac{2(2g_4+g_3)}{3} \pi_{\beta(2)}
\rho^{\alpha\beta(2)} ] \\
 && + \zeta_\alpha [ \frac{4(2g_1+g_2)}{3} B^{\alpha\beta} 
D_\beta{}^\gamma \rho_\gamma +  \frac{4(g_3-4g_4)}{3}
\pi^{\alpha\beta} D_\beta{}^\gamma \rho_\gamma ] \\
 && + \zeta_\alpha (2Mg_1 \pi^{\alpha(2)} - 4Mg_4 B^{\alpha(2)})
\rho_\alpha ].
\end{eqnarray*}
Now we put
\begin{equation}
g_2 = g_1, \qquad g_4 = - \frac{g_3}{2}
\end{equation}
and obtain
$$
\delta_0 {\cal L}_1 + \delta_1 {\cal L}_0 = 
\zeta_\alpha [ (4g_1 B^{\alpha\beta} + 4g_3 \i^{\alpha\beta})
D_\beta{}^\gamma \rho_\gamma + (4Mg_3 B^{\alpha\beta} + 4Mg_1
\pi^{\alpha\beta}) \rho_\beta ].
$$
At last we introduce supertransformations for the spinor
\begin{equation}
\delta \rho^\alpha = 4g_1 B^{\alpha\beta} \zeta_\beta + 4g_3
\pi^{\alpha\beta} \zeta_\beta \nonumber
\end{equation}
and this leaves us with
$$
\delta_0 {\cal L}_1 + \delta_1 {\cal L}_0 = 2 \zeta_\alpha
[ (Mg_3 - M_0g_1) B^{\alpha\beta} + (Mg_1 - M_0g_3) \pi^{\alpha\beta}
] \rho_\beta.
$$
Thus we obtain two solutions
\begin{equation}
M_0 = \pm M, \qquad g_3 = \pm g_1
\end{equation}
Recall that in three dimensions the massive spin $s \ge 1$ boson has
two physical degrees of freedom corresponding to two helicities $\pm
s$. while the massive fermion has only one physical degree of freedom
corresponding to helicity $+ s$ or $-s$. Moreover, the sign of the
helicity is correlated with the sign of the mass term in the free
Lagrangian. 

Combining all things together, we obtain the cubic vertex
\begin{eqnarray}
{\cal L}_1 &=& \Psi_\alpha [ g_1 (E^\alpha{}_\beta B^\beta{}_\gamma -
B^\alpha{}_\beta E^\beta{}_\gamma) \rho^\gamma + g_1 DA \rho^\alpha
\nonumber \\
 && \quad + g_3 (E\pi) \rho^\alpha - \frac{g_3}{2} e^{\alpha\beta} 
(D \varphi - 2M A) \rho_\beta  ], 
\end{eqnarray}
which is invariant under the following supertransformations:
\begin{eqnarray}
\delta B^{\alpha(2)} &=& - 2g_1 \zeta_\beta D^{\alpha(2)} \rho^\beta 
+ 2Mg_3 \zeta^\alpha \rho^\alpha, \qquad
\delta A_{\alpha(2)} = \frac{g_1}{2} \zeta_\alpha \rho_\alpha,
\nonumber \\
\delta \pi_{\alpha(2)} &=&  - g_3 (\zeta_\alpha D^\beta{}_\alpha
\rho_\beta + \zeta_\beta D^\beta{}_\alpha \rho_\alpha), \qquad
\delta \varphi = 2g_3 \zeta_\alpha \rho^\alpha \\
\delta \rho^\alpha &=& 4g_1 B^{\alpha\beta} \zeta_\beta + 4g_3
\pi^{\alpha\beta} \zeta_\beta. \nonumber
\end{eqnarray}

\subsection{Superspin $Y=1$}

In our last example we also begin with the sum of the free Lagrangians
for massive spin-3/2 and spin-1:
\begin{eqnarray}
{\cal L}_0 &=& - \frac{1}{2} \Phi_\alpha D \Phi^\alpha + \frac{1}{2}
\rho_\alpha E^\alpha{}_\beta D \rho^\beta - \frac{M_0}{2} \Phi_\alpha
e^\alpha{}_\beta \Phi^\beta + 2M_0 \Phi_\alpha E^\alpha{}_\beta
\rho^\beta - \frac{3M_0}{2} E \rho_\alpha \rho^\alpha \nonumber \\
 && + E [ \frac{1}{2} B_{\alpha(2)} B^{\alpha(2)} + 4
B^{\alpha\beta} D_\alpha{}^\gamma A_{\beta\gamma} - \frac{1}{2}
\pi_{\alpha(2)} \pi^{\alpha(2)} - \pi^{\alpha(2)} (D_{\alpha(2)}
\varphi - 2M A_{\alpha(2)}) ]
\end{eqnarray}
with the corresponding free equations
\begin{eqnarray}
0 &=& - D \Phi^\alpha - M_0 e^\alpha{}_\beta \Phi^\beta + 2M_0 
E^\alpha{}_\beta \rho^\beta, \nonumber \\
0 &=& E^\alpha{}_\beta D \rho^\beta - 2M_0 E^\alpha{}_\beta \Phi^\beta
- 3M_0  E \rho^\alpha \\
 &=& E [ 2 D^\alpha{}_\beta \rho^\beta + 6M_0 \phi^\alpha - 3M_0
\rho^\alpha ]. \nonumber
\end{eqnarray}
\begin{eqnarray}
0 &=& B_{\alpha(2)} + 2 D_\alpha{}^\beta A_{\alpha\beta}, \qquad
0 = 2 D_\alpha{}^\beta B_{\alpha\beta} + 2M \pi_{\alpha(2)}, \nonumber
\\
0 &=& - \pi_{\alpha(2)} - D_{\alpha(2)} \varphi + 2M A_{\alpha(2)},
\qquad 0 = (D\pi). 
\end{eqnarray}
In this case we begin with the following ansatz for the cubic terms
with one derivative:
\begin{eqnarray}
{\cal L}_{11} &=& g_0 \Psi_\alpha (eB) \Phi^\alpha + g_1 \Psi_\alpha
(E^\alpha{}_\beta B^\beta{}_\gamma - B^\alpha{}_\beta 
E^\beta{}_\gamma) \rho^\gamma + g_2 \Psi_\alpha DA \rho^\alpha
\nonumber \\
 && + g_3 \Psi_\alpha (E\pi) \rho^\alpha + g_4 \Psi_\alpha
e^{\alpha\beta} D \varphi \rho_\beta. \label{s1l1}
\end{eqnarray}
The only new term is the first one with the coefficient $g_0$ so
calculations are similar to the previous case. We need the following
supertransformations:
\begin{eqnarray}
\delta_1 \Phi^\alpha &=& g_0 (eB) \zeta^\alpha, \qquad
\delta_1 \rho^\alpha = (2g_1 B^{\alpha\beta} + 2g_3 \pi^{\alpha\beta})
\zeta_\beta, \nonumber \\
\delta_1 B^{\alpha(2)} &=& - 2g_2 \zeta_\beta D^{\alpha(2)}
\rho^\beta, \qquad \delta_1 A_{\alpha(2)} = 2g_0 \zeta_\beta 
\phi^\beta{}_{\alpha(2)} - g_0 \zeta_\alpha \phi_\alpha + 
\frac{g_1}{2} \zeta_\alpha \rho_\alpha, \\
\delta_1 \pi_{\alpha(2)} &=& 2g_4 (\zeta_\alpha D^\beta{}_\alpha
\rho_\beta + \zeta_\beta D^\beta{}_\alpha \rho_\alpha), \qquad
\delta_1 \varphi = 2g_3 \zeta_\alpha \rho^\alpha. \nonumber
\end{eqnarray}
As in the previous case we have to put
\begin{equation}
g_1 = g_1, \qquad g_4 = - \frac{g_3}{2}
\end{equation}
and this leaves us with
\begin{eqnarray*}
\delta_0 {\cal L}_{11} + \delta_1 {\cal L}_0 &=& \zeta_\alpha
[ 4M_0g_0 B_{\beta(2)} \phi^{\alpha\beta(2)} + (6M_0g_1+4M_0g_0)
B^{\alpha(2)} \phi_\alpha - (3M_0g_1 + 2M_0g_0) B^{\alpha(2)}
\rho_\alpha ] \\
 && + \zeta_\alpha [ - 4Mg_0 \pi_{\beta(2)} \phi^{\alpha\beta(2)}
+ (2Mg_0 + 6M_0g_3) \pi^{\alpha(2)} \phi_\alpha + (2Mg_1 - 3M_0g_3) 
\pi^{\alpha(2)} \rho_\alpha ] \\
 && - 2Mg_3 \zeta_\alpha (A^{\alpha(2)} D_\alpha{}^\beta \rho_\beta
+ A^{\beta(2)} D^\alpha{}_\beta \rho_\beta). 
\end{eqnarray*}
The structure of the cubic terms without derivatives is also
completely determined by the gauge transformations of the vector field
\begin{equation}
{\cal L}_{10} = \Psi_\alpha [ f_0 A \Phi^\alpha - 2Mg_4
e^{\alpha\beta} A \rho_\beta - \frac{f_0}{2} e^\alpha{}_\beta \varphi
\Phi^\beta + f_0 E^\alpha{}_\beta \varphi \rho^\beta] \label{s1l0}
\end{equation}
but this time the invariance is non-trivial and requires
\begin{equation}
\delta \Psi^\alpha = f_0 \Phi^\alpha \xi, \qquad
\delta \Phi^\alpha = - f_0 \Psi^\alpha \xi. 
\end{equation}
At this step we have to introduce additional terms to
supertransformations:
\begin{eqnarray}
\delta_1 \Phi^\alpha &=& f_0 A \zeta^\alpha - \frac{f_0}{2} 
e^\alpha{}_\beta \varphi \zeta^\beta, \qquad \delta_1 \rho^\alpha =
f_0 \varphi \rho^\alpha, \nonumber \\
\delta_1 B^{\alpha(2)} &=& f_0 (2 \zeta_\beta \phi^{\alpha(2)\beta} -
\zeta^\alpha \pi^\alpha) + 2Mg_3 \zeta^\alpha \rho^\alpha, \\
\delta_1 \pi^{\alpha(2)} &=& - 2f_0 (\zeta_\beta \phi^{\alpha(2)\beta}
+ \zeta^\alpha \phi^\alpha) + f_0 \zeta^\alpha \rho^\alpha,  \nonumber
\end{eqnarray}
where
\begin{equation}
f_0 = 2M_0g_0, \qquad g_0 = - g_1
\end{equation}
In this case we also obtain two solutions corresponding to
\begin{equation}
Mg_1 = M_0g_3, \qquad M_0g_1 = Mg_3 \quad \Rightarrow \quad
M_0 = \pm M, \qquad g_3 = \pm g_1
\end{equation}

To be consistent this vertex (\ref{s1l1})+(\ref{s1l0}) must also be
invariant under the (spontaneously broken) $\Phi^\alpha$
supertransformations
\begin{equation}
\delta \Phi^\alpha = D \xi^\alpha + M_0 e^\alpha{}_\beta \xi^\beta,
\qquad \delta \rho^\alpha = 2M_0 \xi^\alpha 
\end{equation}
and this serves as a non-trivial check for the whole construction.
Indeed, by direct calculations (similar to the ones described above),
one can check that the vertex is invariant with the following
corrections:
\begin{eqnarray}
\delta \Psi^\alpha &=& g_0 (eB) \xi^\alpha + f_0 A \xi^\alpha -
\frac{f_0}{2} e^\alpha{}_\beta \varphi \xi^\alpha, \nonumber \\
\delta A^{\alpha(2)} &=& 2g_0 (2\xi_\beta \psi^{\alpha(2)\beta}
- \xi^\alpha \psi^\alpha).  
\end{eqnarray}

\section{Arbitrary superspin $Y > 1$}

In the frame-like multispinor formalism the cubic vertices for the
higher spin bosons and fermions have very simple and regular
structure \cite{Zin21}. So we are ready to consider massive
supermultiplet with arbitrary superspin containing massive boson with
spin $s \ge 2$ and massive fermion with spin $s \pm 1/2$. Taking into
account the experience gained on the lower superspin examples, we
consider the following ansatz for the cubic vertex:
\begin{eqnarray}
{\cal L}_1 &=& \sum_k (-1)^{k+1} \Psi_\alpha 
[ h_{1,k} \Omega_{\beta(2k)} \Phi^{\alpha\beta(2k)}
+ h_{2,k} \Omega^{\alpha\beta(2k-1)} \Phi_{\beta(2k-1)} \nonumber \\
 && \qquad \qquad + h_{3,k} H_{\beta(2k)} \Phi^{\alpha\beta(2k)}
+ h_{4,k} H^{\alpha\beta(2k-1)} \Phi_{\beta(2k-1)} ] \nonumber \\
 && + \Psi_\alpha [ g_0 (eB) \Phi^\alpha + g_1 (E^\alpha{}_\beta
B^\beta{}_\gamma - B^\alpha{}_\beta E^\beta{}_\gamma) \rho^\gamma
+ g_2 DA \rho^\alpha + g_3 (E\pi) \rho^\alpha + g_4
e^{\alpha\beta} D \varphi \rho_\beta \nonumber \\
 && \quad + f_0 A \Phi^\alpha + f_1 e^{\alpha\beta} A \rho_\beta
+ f_2 e^\alpha{}_\beta \varphi \Phi^\beta + f_3 E^{\alpha\beta}
\varphi \rho_\beta + f_4 (eH) \rho^\alpha ]. 
\end{eqnarray}
To simplify presentation we move calculation details into Appendix and
present here only final results. First of all, the supersymmetry
completely fixes all the unknown coefficients in terms of just one
arbitrary coupling constant $h_0$:
\begin{eqnarray}
h_{1,k}  &=& \sqrt{k(s \pm (k+1))}h_0, \qquad
h_{2,k} = \sqrt{\frac{2k(k+1)(s \mp k)}{(2k+1)}}h_0, \nonumber \\
h_{3,k} &=& \frac{Ms}{2(k+1)} \sqrt{\frac{(s \pm (k+1))}{k}}h_0,
\qquad h_{4,k} = Ms\sqrt{\frac{(s \mp k)}{2k(k+1)(2k+1)}}h_0,
\label{sol} \\
g_0 &=& - \frac{1}{2}\sqrt{(s \pm 1)}h_0, \qquad
g_1 = g_2 = g_3 = 2g_4 = \sqrt{\frac{s}{2}}h_0, \qquad
f_0 = 2Msg_0, \nonumber \\
f_1 &=& Msg_1, \qquad f_2 = Msg_0, \qquad
f_3 = - M(s \pm 1)g_1, \qquad f_4 = \frac{a_0}{4}g_1. \nonumber
\end{eqnarray}
Here upper/lower signs correspond to integer/half-integer superspin. 
The invariance under the local supertransformations requires
non-trivial transformations for all components. For the bosonic
one-forms we obtain
\begin{eqnarray}
\delta \Omega^{\alpha(2k)} &=& h_{3,k} \zeta_\beta 
\Phi^{\alpha(2k)\beta} + \frac{h_{4,k}}{2k} \zeta^\alpha
\Phi^{\alpha(2k-1)}, \nonumber \\
\delta H^{\alpha(2k)} &=& h_{1,k} \zeta_\beta 
\Phi^{\alpha(2k)\beta} + \frac{h_{2,k}}{2k} \zeta^\alpha
\Phi^{\alpha(2k-1)}, \nonumber \\
\delta \Omega^{\alpha(2)} &=& h_{3,1} \zeta_\beta 
\Phi^{\alpha(2)\beta} + \frac{h_{4,1}}{2} \zeta^\alpha \Phi^\alpha
- f_4 e^{\alpha(2)} \zeta_\beta \rho^\beta, \\
\delta H^{\alpha(2)} &=& h_{1,1} \zeta_\beta \Phi^{\alpha(2)\beta} +
\frac{h_{2,1}}{2} \zeta^\alpha \Phi^\alpha, \nonumber \\
\delta A &=& - g_0 \zeta_\alpha \Phi^\alpha + \frac{g_1}{2}
\zeta_\alpha e^{\alpha\beta} \rho_\beta, \nonumber
\end{eqnarray}
while for the bosonic zero-forms we get
\begin{eqnarray}
\delta B^{\alpha(2)} &=&  - g_2 \zeta_\beta D^{\alpha(2)} \rho^\beta  
+ f_0 (\zeta_\beta \phi^{\alpha(2)\beta} + \zeta^\alpha \phi^\alpha) 
 + f_1 \zeta^\alpha \rho^\alpha, \nonumber \\
\delta \pi^{\alpha(2)} &=& - g_4 (\zeta^\alpha D^\alpha{}_\beta
\rho^\beta + \zeta_\beta D^{\alpha\beta} \rho^\alpha) 
- 2f_2 (\zeta_\beta \phi^{\alpha(2)\beta} + \zeta^\alpha \phi^\alpha)
+ \frac{f_3}{2} \zeta^\alpha \rho^\alpha, \\
\delta \varphi &=&  - g_3 \zeta_\alpha \rho^\alpha. \nonumber
\end{eqnarray}
Recall that the zero-forms $\phi^{\alpha(3)}$ and $\phi^\alpha$ come
from the decomposition
$$
\Phi^\alpha = e_{\beta(2)} \phi^{\alpha\beta(2)} + e^{\alpha(2)}
\phi^\alpha.
$$
At the same time the transformations for the fermions look like:
\begin{eqnarray}
\delta \Phi^{\alpha(2k+1)} &=& \frac{h_{1,k}}{(2k+1)}
\Omega^{\alpha(2k)} \zeta^\alpha + h_{2,k+1} 
\Omega^{\alpha(2k+1)\beta} \zeta_\beta \nonumber \\
 && + \frac{h_{3,k}}{(2k+1)} H^{\alpha(2k)} \zeta^\alpha + h_{4,k+1} 
H^{\alpha(2k+1)\beta} \zeta_\beta, \nonumber \\
\delta \Phi^\alpha &=& h_{2,1} \Omega^{\alpha\beta} \zeta_\beta +
h_{4,1} H^{\alpha\beta} \zeta_\beta  \\
 && + g_0 (eB) \zeta^\alpha + F_0 A \zeta^\alpha + f_2
e^\alpha{}_\beta \varphi \zeta^\beta, \nonumber \\
\delta \rho^\alpha &=& 2g_1 B^{\alpha\beta} \zeta_\beta + g_1
\pi^{\alpha\beta} \zeta_\beta - f_3 \varphi \zeta^\alpha. \nonumber
\end{eqnarray}
Thus supersymmetry completely fixes both the structure of the cubic
vertex as well as the supertransformations for all components.
However, the whole construction is based on the gauge invariant
description of massive higher spin bosons and fermions. So we have to
check that our cubic vertex is consistent with all these gauge
symmetries.

\noindent
{\bf Fermionic gauge transformations} Here the gauge invariance is
achieved with the following corrections for the gravitino
\begin{eqnarray}
\delta \Psi^\alpha &=& \sum_{k=1} (-1)^{k+1} [ - h_{1,k}
\Omega_{\beta(2k)} \zeta^{\alpha\beta(2k)} + h_{2,k+1}
\Omega^{\alpha\beta(2k+1)} \zeta_{\beta(2k+1)} \nonumber \\
 && \qquad - h_{3,k} H_{\beta(2k)} \zeta^{\alpha\beta(2k)}
+ h_{4,k+1} H^{\alpha\beta(2k+1)} \zeta_{\beta(2k+1)} ] \nonumber \\
 && - h_{2,1} \Omega^{\alpha\beta} \zeta_\beta - h_{4,1}
H^{\alpha\beta} \zeta_\beta - g_0 (eB) \zeta^\alpha
- f_0 A \zeta^\alpha + f_2 e^{\alpha\beta} \varphi \zeta_\beta,
\end{eqnarray}
as well as for all bosonic components
\begin{eqnarray}
\delta \Omega^{\alpha(2k)} &=& - h_{3,k} \Psi_\beta
\zeta^{\alpha(2k)\beta} - \frac{h_{4,k}}{2k} \Psi^\alpha
\zeta^{\alpha(2k-1)}, \nonumber \\
\delta H^{\alpha(2k)} &=& - h_{1,k} \Psi_\beta
\zeta^{\alpha(2k)\beta} - \frac{h_{2,k}}{2k} \Psi^\alpha
\zeta^{\alpha(2k-1)}, \\
\delta A &=& g_0 \Psi_\alpha \zeta^\alpha. \nonumber
\end{eqnarray}

\noindent
{\bf Bosonic gauge transformations} In this case the gravitino must
transform as follows
\begin{eqnarray}
\delta \Psi^\alpha &=& \sum_{k=1} (-1)^{k+1} [
h_{1,k}\Phi^{\alpha\beta(2k)} \eta_{\beta(2k)} + h_{2,k}
\eta^{\alpha\beta(2k-1)} \Phi_{\beta(2k-1)} \nonumber \\
 && \qquad + h_{3,k} \Phi^{\alpha\beta(2k)} \xi_{\beta(2k)} + h_{4,k}
\xi^{\alpha\beta(2k-1)} \Phi_{\beta(2k-1)} ] 
 + f_0 \Phi^\alpha \xi, 
\end{eqnarray}
while corrections for the fermionic components look like:
\begin{eqnarray}
\delta \Phi^{\alpha(2k+1)} &=& - \frac{h_{1,k}}{(2k+1)}
\Psi^\alpha \eta^{\alpha(2k)} - h_{2,k+1} \Psi_\beta
\eta^{\alpha(2k+1)\beta} \nonumber \\
 && - \frac{h_{3,k}}{(2k+1)} \Psi^\alpha \xi^{\alpha(2k)} - h_{4,k+1}
\Psi_\beta \xi^{\alpha(2k+1)\beta}, \\
\delta \Phi^\alpha &=& - h_{2,1} \Psi_\beta \eta^{\alpha\beta}
- h_{4,1} \Psi_\beta \xi^{\alpha\beta} - f_0 \Psi^\alpha \xi.
\nonumber
\end{eqnarray}
Thus the cubic vertex constructed is completely consistent with all
the gauge symmetries.

\appendix

\section{Calculation details}

The lower spin part of the cubic vertex is already familiar from the
lower superspin examples, so let us restrict ourselves with higher
spin part of the vertex:
\begin{eqnarray}
\Delta {\cal L}_1 &=& \sum_k (-1)^{k+1} \Psi_\alpha 
[ h_{1,k} \Omega_{\beta(2k)} \Phi^{\alpha\beta(2k)}
+ h_{2,k} \Omega^{\alpha\beta(2k-1)} \Phi_{\beta(2k-1)} \nonumber \\
 && \qquad + h_{3,k} H_{\beta(2k)} \Phi^{\alpha\beta(2k)}
+ h_{4,k} H^{\alpha\beta(2k-1)} \Phi_{\beta(2k-1)} ]. 
\end{eqnarray}
Taking the variation of this part under the local supertransformations
we obtain:
\begin{eqnarray}
\delta (\Delta {\cal L}_1) &=& \sum_k (-1)^{k+1} \zeta_\alpha [
h_{1,k} (- D \Omega_{\beta(2k)} \Phi^{\alpha\beta(2k)}
+ \Omega_{\beta(2k)} D \Phi^{\alpha\beta(2k)}) \nonumber \\
 && \qquad + h_{2,k} (- D \Omega^{\alpha\beta(2k-1)}
\Phi_{\beta(2k-1)} + \Omega^{\alpha\beta(2k-1)} D \Phi_{\beta(2k-1)}
\nonumber \\
 && \qquad [ h_{3,k} (- D H_{\beta(2k)} \Phi^{\alpha\beta(2k)}
+ H_{\beta(2k)} D \Phi^{\alpha\beta(2k)}) \nonumber \\
 && \qquad + h_{4,k} (- D H^{\alpha\beta(2k-1)} \Phi_{\beta(2k-1)}
+ H^{\alpha\beta(2k-1)} D \Phi_{\beta(2k-1)} ].
\end{eqnarray}
From this expression it is easy to read out all necessary corrections
to the supertransformations of the bosonic and fermionic fields:
\begin{eqnarray}
\delta \Omega^{\alpha(2k)} &=& h_{3,k} \zeta_\beta 
\Phi^{\alpha(2k)\beta} + \frac{h_{4,k}}{2k} \zeta^\alpha
\Phi^{\alpha(2k-1)}, \nonumber \\
\delta H^{\alpha(2k)} &=& h_{1,k} \zeta_\beta 
\Phi^{\alpha(2k)\beta} + \frac{h_{2,k}}{2k} \zeta^\alpha
\Phi^{\alpha(2k-1)}, 
\end{eqnarray}
\begin{eqnarray}
\delta \Phi^{\alpha(2k+1)} &=& \frac{h_{1,k}}{(2k+1)}
\Omega^{\alpha(2k)} \zeta^\alpha + h_{2,k+1} 
\Omega^{\alpha(2k+1)\beta} \zeta_\beta \nonumber \\
 && + \frac{h_{3,k}}{(2k+1)} H^{\alpha(2k)} \zeta^\alpha + h_{4,k+1} 
H^{\alpha(2k+1)\beta} \zeta_\beta.  
\end{eqnarray}
These corrections produce a lot of terms proportional to the free
bosonic (\ref{bos_eq}) and fermionic (\ref{ferm_eq}) equations. As a
result, all the terms with explicit derivatives cancel leaving us
with the terms without derivatives. As an illustration let us consider
all terms with $\Omega^{\alpha(2k)}$ and organize them by the
fermionic fields they contain. \\
Terms with $\Phi^{\alpha(2k+3)}$:
$$
0 = [ - a_kh_{1,k+1} + c_{k+1}h_{1,k}] \zeta_\alpha
e_{\gamma(2)} \Omega_{\beta(2k)} \Phi^{\alpha\beta(2k)\gamma(2)}.
$$
This gives us
\begin{equation}
a_kh_{1,k+1} = c_{k+1}h_{1,k}. \label{eq1}
\end{equation}
Terms with $\Phi^{\alpha(2k-3)}$
$$
0 = [ - \frac{(k+1)}{(k-1)}a_{k-1}h_{2,k-1} + c_{k-1}h_{2,k}]
\zeta_\alpha e_{\gamma(2)} \Omega^{\alpha\beta(2k-3)\gamma(2)}
\Phi_{\beta(2k-3)}.
$$
This gives
\begin{equation}
(k+1)a_{k-1}h_{2,k-1} = (k-1)c_{k-1}h_{2,k}.
\end{equation}
Terms with $\Phi^{\alpha(2k+1)}$:
\begin{eqnarray*}
0 &=& [ \frac{d_k}{(2k+1)}h_{1,k} + \frac{a_k}{(k+1)}h_{2,k+1}]
\zeta_\alpha e^\alpha{}_\gamma \Omega_{\beta(2k)} 
\Phi^{\beta(2k)\gamma} \\
 && + [ \frac{2kd_k}{(2k+1)}h_{1,k} - 2kh_{3,k}] \zeta_\alpha
e^\gamma{}_\delta \Omega_{\beta(2k-1)\gamma} 
\Phi^{\alpha\beta(2k-1)\delta} \\
 && + [ \frac{ka_k}{(k+1)}h_{2,k+1} - c_kh_{2,k}] \zeta_\alpha 
e_{\gamma(2)} \Omega^\alpha{}_{\beta(2k-1)} 
\Phi^{\beta(2k-1)\gamma(2)}.
\end{eqnarray*}
In this case we have to take into account an identity
$$
e^\alpha{}_\gamma \Omega_{\beta(2k)} 
\Phi^{\beta(2k)} - e^\delta{}_\gamma
\Omega_{\beta(2k-1)\delta} \Phi^{\alpha\beta(2k-1)\gamma} = 
e_{\gamma(2)} \Omega^\alpha{}_{\beta(2k-1)}
\Phi^{\beta(2k-1)\gamma(2)}, 
$$
so we obtain only two independent equations:
\begin{eqnarray}
0 &=&  \frac{d_k}{(2k+1)}h_{1,k} + a_kh_{2,k+1} - c_kh_{2,k}, \\
0 &=& \frac{2kd_k}{(2k+1)}h_{1,k} - 2kh_{3,k}
- \frac{ka_k}{(k+1)}h_{2,k+1} + c_kh_{2,k}.
\end{eqnarray}
Terms with $\Phi^{\alpha(2k-1)}$
\begin{eqnarray*}
0 &=& [ - \frac{(k+1)a_{k-1}}{(k-1)}h_{1,k-1} 
+ \frac{(2k-1)c_k}{(2k+1)}h_{1,k}] \zeta_\alpha
e^{\gamma(2)} \Omega_{\beta(2k-2)\gamma(2)} \Phi^{\alpha\beta(2k-2)} 
\\
 && + [  \frac{2c_k}{(2k+1)}h_{1,k} + h_{4,k}] \zeta_\alpha
e^{\alpha\gamma} \Omega_{\beta(2k-1)\gamma} \Phi^{\beta(2k-1)} \\
 && + [ - d_{k-1}h_{2,k} + (2k-1)h_{4,k}] \zeta_\alpha 
e^\gamma{}_\delta \Omega^\alpha{}_{\beta(2k-2)\gamma}
\Phi^{\beta(2k-2)\delta}.
\end{eqnarray*}
Again, using an identity
$$
e^{\alpha\gamma} \Omega_{\beta(2k-1)\gamma}
\Phi^{\beta(2k-1)} - e^{\gamma(2)}
\Omega_{\beta(2k-2)\gamma(2)} \Phi^{\alpha\beta(2k-2)}
= e^\gamma{}_\delta \Omega^\alpha{}_{\beta(2k-2)\gamma}
\Phi^{\beta(2k-2)\delta}, 
$$
we obtain the last pair of equations
\begin{eqnarray}
0 &=& - \frac{(k+1)a_{k-1}}{(k-1)}h_{1,k-1} 
+ c_kh_{1,k} + h_{4,k}, \\
0 &=& - d_{k-1}h_{2,k} + 2kh_{4,k}
+ \frac{2c_k}{(2k+1)}h_{1,k}. \label{eq2}
\end{eqnarray}
The equations (\ref{eq1}) - (\ref{eq2}) appear to be consistent and
give the solution (\ref{sol}). 

Similarly, let us consider terms with the field $H^{\alpha(2k)}$. \\
Terms with $\Phi^{\alpha(2k+3)}$
$$
0 = [ - \frac{(k+2)a_k}{k}h_{3,k+1} + c_{k+1}h_{3,k}] \zeta_\alpha
e_{\gamma(2)} H_{\beta(2k)} \Phi^{\alpha\beta(2k)\gamma(2)}. 
$$
Thus
\begin{equation}
(k+2)a_kh_{3,k+1} = kc_{k+1}h_{3,k}. \label{eq3}
\end{equation}
Terms with $\Phi^{\alpha(2k-3)}$
$$
0 = [ - a_{k-1}h_{4,k-1} + c_{k-1}h_{4,k}] \zeta_\alpha
e_{\gamma(2)} H^{\alpha\beta(2k-3)\gamma(2)} \Phi_{\beta(2k-3)}.
$$
Thus
\begin{equation}
a_{k-1}h_{4,k-1} = c_{k-1}h_{4,k}.
\end{equation}
Terms with $\Phi^{\alpha(2k+1)}$
\begin{eqnarray*}
0 &=& [ - 2b_kh_{1,k} + \frac{2kd_k}{(2k+1)}h_{3,k}] \zeta_\alpha
e^\gamma{}_\delta H_{\beta(2k-1)\gamma} 
\Phi^{\alpha\beta(2k-1)\delta} \\
 && + [\frac{d_k}{(2k+1)}h_{3,k} + \frac{(k+2)a_k}{k(k+1)}h_{4,k+1}]
\zeta_\alpha e^\alpha{}_\gamma H_{\beta(2k)} \Phi^{\beta(2k)\gamma} \\
 && + [ \frac{(k+2)a_k}{(k+1)}h_{4,k+1} - c_kh_{4,k}] \zeta_\alpha
e_{\gamma(2)} H^\alpha{}_{\beta(2k-1)} \Phi^{\beta(2k-1)\gamma(2)}.
\end{eqnarray*}
Using the appropriate identity we obtain
\begin{eqnarray}
0 &=& - 2b_kh_{1,k} + d_kh_{3,k} + \frac{(k+2)a_k}{k(k+1)}h_{4,k+1}, 
\\
0 &=& \frac{(k+2)a_k}{k}h_{4,k+1} - c_kh_{4,k}
+ \frac{d_k}{(2k+1)}h_{3,k}. 
\end{eqnarray}
Terms with $\Phi^{\alpha(2k-1)}$
\begin{eqnarray*}
0 &=& [ \frac{b_k}{k}h_{2,k} + \frac{2c_k}{(2k+1)}h_{3,k}]
\zeta_\alpha e^\alpha{}_\gamma H^{\beta(2k-1)\gamma}
\Phi_{\beta(2k-1)} \\
 && + [ \frac{(2k-1)b_k}{k}h_{2,k} - d_{k-1}h_{4,k}] \zeta_\alpha
e_\gamma{}^\delta H^{\alpha\beta(2k-2)\gamma} 
\Phi_{\beta(2k-2)\delta} \\
 && + [ - a_{k-1}h_{3,k-1} + \frac{(2k-1)c_k}{(2k+1)}h_{3,k}]
\zeta_\alpha e_{\gamma(2)} H^{\beta(2k-2)\gamma(2)} 
\Phi^\alpha{}_{\beta(2k-2)}. 
\end{eqnarray*}
Here we obtain\begin{eqnarray}
0 &=& 2b_kh_{2,k} - d_{k-1}h_{4,k} + \frac{2c_k}{(2k+1)}h_{3,k}, \\
0 &=& - a_{k-1}h_{3,k-1} + c_kh_{3,k} + \frac{b_k}{k}h_{2,k}.
\label{eq4}
\end{eqnarray}
Equations (\ref{eq3}) - (\ref{eq4}) are also consistent and give
exactly the same solution (\ref{sol}).

\end{document}